\documentclass[%
 reprint,
superscriptaddress,
 amsmath,amssymb,
 aps,
longbibliography
]{revtex4-1}

\usepackage{aas_macros}
\usepackage{graphicx}
\usepackage{dcolumn}
\usepackage{bm}
\usepackage{hyperref}
\usepackage{color}
\usepackage[mathlines]{lineno}
\usepackage{tabularx}
\usepackage{subfigure}
\usepackage[normalem]{ulem}
\usepackage[export]{adjustbox}
\usepackage{url}
\usepackage{lineno}
\usepackage[flushleft]{threeparttable}
\usepackage{makecell}
\usepackage{multirow}

\hypersetup{
    pdfnewwindow=true,    
    colorlinks=true,      
    linkcolor=blue,       
    citecolor=blue,       
    filecolor=blue,       
    urlcolor=blue         
}

\newcommand{\ga}{$\gamma$}

\newcolumntype{Y}{>{\centering\arraybackslash}X}
\def\orcid#1{\href{https://orcid.org/#1}{\includegraphics[keepaspectratio,width=1.1em]{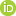}}}

\usepackage{natbib}

\newcommand{\INFN}{INFN - Sezione di Napoli, Complesso Univ. Monte S. Angelo, I-80126 Napoli, Italy}
\newcommand{\UNINA}{Dipartimento di Fisica ``Ettore Pancini'', Università degli studi di Napoli ``Federico II'', Complesso Univ. Monte S. Angelo, I-80126 Napoli, Italy}
\newcommand{\SSM}{Scuola Superiore Meridionale, Università degli studi di Napoli ``Federico II'', Largo San Marcellino 10, 80138 Napoli, Italy}
\newcommand{\NBIA}{Niels Bohr International Academy, Niels Bohr Institute, University of Copenhagen, Copenhagen, Denmark}

\begin{document}
\title{Constraints on Heavy Decaying Dark Matter from 570 Days of LHAASO Observations}


\author{Zhen Cao}
\affiliation{Key Laboratory of Particle Astrophyics \& Experimental Physics Division \& Computing Center, Institute of High Energy Physics, Chinese Academy of Sciences, 100049 Beijing, China}
\affiliation{University of Chinese Academy of Sciences, 100049 Beijing, China}
\affiliation{TIANFU Cosmic Ray Research Center, Chengdu, Sichuan,  China}

\author{F. Aharonian}
\affiliation{Dublin Institute for Advanced Studies, 31 Fitzwilliam Place, 2 Dublin, Ireland }
\affiliation{Max-Planck-Institut for Nuclear Physics, P.O. Box 103980, 69029  Heidelberg, Germany}

\author{Q. An}
\affiliation{State Key Laboratory of Particle Detection and Electronics, China}
\affiliation{University of Science and Technology of China, 230026 Hefei, Anhui, China}

\author{Axikegu}
\affiliation{School of Physical Science and Technology \&  School of Information Science and Technology, Southwest Jiaotong University, 610031 Chengdu, Sichuan, China}

\author{L.X. Bai}
\affiliation{College of Physics, Sichuan University, 610065 Chengdu, Sichuan, China}

\author{Y.X. Bai}
\affiliation{Key Laboratory of Particle Astrophyics \& Experimental Physics Division \& Computing Center, Institute of High Energy Physics, Chinese Academy of Sciences, 100049 Beijing, China}
\affiliation{TIANFU Cosmic Ray Research Center, Chengdu, Sichuan,  China}

\author{Y.W. Bao}
\affiliation{School of Astronomy and Space Science, Nanjing University, 210023 Nanjing, Jiangsu, China}

\author{D. Bastieri}
\affiliation{Center for Astrophysics, Guangzhou University, 510006 Guangzhou, Guangdong, China}

\author{X.J. Bi}
\affiliation{Key Laboratory of Particle Astrophyics \& Experimental Physics Division \& Computing Center, Institute of High Energy Physics, Chinese Academy of Sciences, 100049 Beijing, China}
\affiliation{University of Chinese Academy of Sciences, 100049 Beijing, China}
\affiliation{TIANFU Cosmic Ray Research Center, Chengdu, Sichuan,  China}

\author{Y.J. Bi}
\affiliation{Key Laboratory of Particle Astrophyics \& Experimental Physics Division \& Computing Center, Institute of High Energy Physics, Chinese Academy of Sciences, 100049 Beijing, China}
\affiliation{TIANFU Cosmic Ray Research Center, Chengdu, Sichuan,  China}

% ...

\collaboration{LHAASO Collaboration}
\email{lizhe@ihep.ac.cn; chensz@ihep.ac.cn}

\author{S.Ando}
\affiliation{GRAPPA Institute, University of Amsterdam, 1098 XH Amsterdam, The Netherlands}
\affiliation{Kavli Institute for the Physics and Mathematics of the Universe~$\left( Kavli~IPMU,~WPI\right)$, University of Tokyo, Kashiwa, Chiba 277-8583, Japan}

\author{M.Chianese~\orcid{0000-0001-8261-4441}}
\email{chianese@na.infn.it}
\thanks{\scriptsize \!\! \href{http://orcid.org/0000-0001-8261-4441}{orcid.org/0000-0001-8261-4441}}
\affiliation{\UNINA}
\affiliation{\INFN}

\author{D.F.G. Fiorillo}
\affiliation{\UNINA}
\affiliation{\INFN}
\affiliation{\NBIA}

\author{G.Miele}
\affiliation{\UNINA}
\affiliation{\INFN}
\affiliation{\SSM}

\author{K.C.Y. Ng~\orcid{0000-0001-8016-2170}}
\email{kcyng@cuhk.edu.hk}
\thanks{\scriptsize \!\! \href{http://orcid.org/0000-0001-8016-2170}{orcid.org/0000-0001-8016-2170}}
\affiliation{Department of Physics, The Chinese University of Hong Kong, Shatin, New Territories, Hong Kong, China}

\date{\today}

\begin{abstract}
The kilometer squarearray~(KM2A) of the large high altitude air shower observatory (LHAASO) aims at surveying the northern $\gamma$-ray sky at energies above 10 TeV with unprecedented sensitivity. \ga-ray observations have long been one of the most powerful tools for dark matter searches, as, e.g., high-energy \ga rays could be produced by the decays of heavy dark matter particles. In this Letter, we present the first dark matter analysis with LHAASO-KM2A, using the first 340~days of data from 1/2-KM2A and 230~days of data from 3/4-KM2A.  Several regions of interest are used to search for a signal and account for the residual cosmic-ray background after \ga/hadron separation. We find no excess of dark matter signals, and thus place some of the strongest $\gamma$-ray constraints on the lifetime of heavy dark matter particles with mass between $10^5$ and $10^9$~GeV. Our results with LHAASO are robust, and have important implications for dark matter interpretations of the diffuse astrophysical high-energy neutrino emission.
\end{abstract}

\maketitle


\textbf{Introduction---}
Dark matter (DM) is one of the cornerstones of fundamental physics and cosmology, as it accounts for most of the mass of the Universe. So far, DM has evaded all the attempts to detect its nongravitational interactions~\cite{Kahlhoefer:2017dnp,Schumann:2019eaa,PerezdelosHeros:2020qyt}; the identification of its nature is one of the primary goals in modern science~\cite{Bertone:2018krk,AlvesBatista:2021gzc}. In this context, DM indirect-detection searches represent a powerful tool that leverages astrophysical data to probe a variety of DM candidates. Among all the astrophysical messengers, high-energy $\gamma$ rays have long been an important avenue for achieving some of the best sensitivities in DM searches~\cite{Fermi-LAT:2012pls,Fermi-LAT:2015att,HESS:2016glm,Fermi-LAT:2016uux,VERITAS:2017tif,Abeysekara:2017jxs,MAGIC:2020ceg}. In this regard, very-high-energy (VHE) $\gamma$ rays offer a unique possibility to probe heavy DM particles with masses above 100~TeV.

In recent years, VHE $\gamma$ rays have been detected from several Galactic sources~\cite{Abramowski:2016mir,Abeysekara:2019gov,Carpet-3:2021omd,LHAASO:2021nature,LHAASO:2021science} as well as from the whole Galactic plane~\cite{Amenomori:2021gmk}. Away from the Galactic plane, upper limits have been placed on the isotropic diffuse $\gamma$-ray flux above 100~TeV~\cite{Chantell:1997gs,Apel:2017ocm,Harding:2019tez}. While the $\gamma$-ray emission from extragalactic sources is significantly suppressed due to the cosmic $\gamma$-ray absorption, detectable high-latitude VHE $\gamma$ rays could be produced through the decays of heavy DM particles in the Galactic halo, as DM annihilations are theoretically disfavored by the unitarity bound in the VHE regime~\cite{Griest:1989wd}. Decaying heavy DM has been theorized in several models, including WIMPIzillas~\cite{Chung:1998zb,Benakli:1998ut,Kolb:1998ki}, glueballs~\cite{Faraggi:2000pv,Boddy:2014yra,Forestell:2016qhc,Halverson:2016nfq}, gravitinos~\cite{Pagels:1981ke,Steffen:2006hw,Ishiwata:2008cu}, frozen-in DM~\cite{Garny:2015sjg,ReFiorentin:2016rzn,Chianese:2016smc,Kolb:2017jvz} and other proposals~\cite{Higaki:2014dwa,Dev:2016qbd,DiBari:2016guw,Contino:2018crt,Babichev:2018mtd,Kim:2019udq,Dudas:2020sbq,Kramer:2020sbb,Hambye:2020lvy,Garcia:2020hyo}. Interestingly, it has also been proposed~\cite{Feldstein:2013kka,Esmaili:2013gha,Chianese:2016opp} as a source of the diffuse TeV-PeV neutrino flux observed by IceCube~\cite{Aartsen:2013jdh,Aartsen:2014gkd,Abbasi:2020jmh}. Such a scenario has been long studied with multimessenger observations~\cite{IceCube:2011kcp,Esmaili:2012us,Murase:2012xs,Esmaili:2014rma,Esmaili:2015xpa,Murase:2015gea,Chianese:2016kpu,Cohen:2016uyg,Bhattacharya:2017jaw,Chianese:2017nwe,Aartsen:2018mxl,Kachelriess:2018rty,Blanco:2018esa,Bhattacharya:2019ucd,Arguelles:2019boy,Chianese:2019kyl,Dekker:2019gpe,Ishiwata:2019aet,Dekker:2019gpe,Viana:2019ucn,Neronov:2020wir,Ng:2020ghe,Esmaili:2021yaw,Maity:2021umk,Chianese:2021htv,Chianese:2021jke, IceCube:2022clp}. Nevertheless, DM contributions to the diffuse high-energy neutrino flux remain a viable possibility~\cite{Chianese:2019kyl}.

The large high altitude air shower observatory (LHAASO)~\cite{Zhen:2019lnn} is a general purpose, continuously operating air shower cosmic-ray and $\gamma$-ray detector located in southwest China, which completed its construction in 2021. It mainly consists of the KM2A~(kilometer square array), WCDA~(water cherenkov detector array), and WFCTA~(wide field of view Cherenkov telescope array). Together, it is sensitive to the $\gamma$-ray sky from 100\,GeV to 1\,PeV, and has for the first time detected PeV \ga rays from astrophysical sources~\cite{LHAASO:2021nature, LHAASO:2021science}.

In this Letter, we utilize the data from partially completed KM2A to search for signatures of DM decays.


\textbf{KM2A Data Analysis---}
KM2A is a ground-based full-duty extensive air shower (EAS) array dedicated to VHE $\gamma$-ray astronomy above 10\,TeV. It has an excellent $\gamma$/hadron separation capability by using both electromagnetic particle detectors (EDs) and underground muon detectors (MDs). The EDs are plastic scintillation detectors and the MDs are water Cherenkov detectors with 2.5m soil overburden~\cite{He:2018hh}. With a large field of view, $\sim2\, \mathrm{sr}$, KM2A covers about $60\%$ of the sky daily~\cite{lhaaso:whitepaper, Aharonian:2020iou}. 

In this work, we consider data from the partially completed KM2A, including 340 days from the 1/2-KM2A~(2365 out of 5216 EDs and 578 out of 1188 MDs, covering an area of 0.432\,km$^{2}$), from December 27, 2019 to November 30, 2020, and 230 days from the 3/4-KM2A~(3978 out of 4901 EDs and 917 out of 1188 MDs, covering an area of 0.727\,km$^{2}$), from December 1, 2020 to July 19, 2021. We employ the same data quality cuts, event selection, and detector simulation as in Ref.~\cite{Aharonian:2020iou} for both 1/2-KM2A and 3/4-KM2A. The angular and energy resolution of the two datasets are similar, with the latter being slightly better. At 100 TeV, the angular and energy resolutions are about 0.3$^\circ$ and 20\%~\cite{Aharonian:2020iou}, respectively.

The field of view of LHAASO covers the celestial northern sky~(Fig.~4 in Ref.~\cite{lhaaso:whitepaper}). Given that the DM signal is expected to be higher with smaller galactocentric radius, and to reduce the potential diffuse astrophysical emission from the northern Fermi bubble and the Galactic plane, we consider one fiducial search region of interest~(${\rm ROI}$), labeled as ${\rm ROI}_0$, around $15^\circ \leq b \leq 45^\circ$ and $30^\circ \leq \ell \leq 60^\circ$. We also consider four control regions~(labeled ${\rm ROI}_1-{\rm ROI}_4$) away from ${\rm ROI}_0$ for the purpose of constraining the isotropic cosmic-ray background. These regions are selected to avoid the Fermi bubbles and the Galactic plane as well. 

Importantly, we also require ${\rm ROI}_1-{\rm ROI}_4$ to have the same declination and angular size (0.274~sr) as ${\rm ROI}_0$. This ensures that all the ROIs have the same detector responses, eliminating potential systematics in the declination dependence of the detector response. Following these criteria, ${\rm ROI}_1-{\rm ROI}_4$ are chosen by shifting ${\rm ROI}_0$ along the RA direction by 90$^\circ$, 135$^\circ$, 240$^\circ$, and 285$^\circ$, respectively. The exposure time for ${\rm ROI}_0$ to ${\rm ROI}_4$ are 523, 510, 523, 527, and 529 days, respectively. Due to being shifted to larger galactocentric radii, the expected DM $\gamma$-ray fluxes from ${\rm ROI}_1-{\rm ROI}_4$ are a factor of a few smaller than the one from ${\rm ROI}_0$. For more details see Supplemental Material~\ref{sec:supple_rois}.

We partition the data from $10^5$ to $10^{6.2}$\,GeV with 6 energy bins in logarithmic space, which are wider than the energy resolution of the detector~\cite{Aharonian:2020iou}. The \ga/hadron separation is then applied by considering the ratio of the detected muons and electrons~[see Eq.~(7) in Ref.~\cite{Aharonian:2020iou}]. To further reduce the background, we adopt a more stringent \ga/hadron cut parameter than the one in point-source analyses~\cite{Aharonian:2020iou}. In this analysis, the \ga-ray survival rate is lowered to be at least 50\% of the injected gamma-ray events in detector simulation, with the cosmic-ray survival rate further lowered down to $1.86 \times 10^{-6}$ around 1~PeV in the observed data (see Supplemental Material~\ref{sec:supple_gp} for details and validation of the cut). Even with such a strong cut, we still expect that most of the residual events are misidentified cosmic-ray events. Table~\ref{table:data} shows the events after \ga/hadron separation. 
\begin{table}[t!]
\caption{Residual events after \ga/hadron separation in the search (${\rm ROI}_0$) and control (${\rm ROI}_1-{\rm ROI}_4$) regions with an observations of 340 days with 1/2-KM2A and 230 days with 3/4-KM2A.}
\begin{center}
\begin{threeparttable}
\begin{tabular}{r r r r r r}
\makecell*[c] {Energy bin \\ $\left[\log_{10} (\frac{E}{\rm GeV}) \right]$ } \vline & $N_{\rm ROI_0}$ \vline & $N_{\rm ROI_1}$ \vline & $N_{\rm ROI_2}$ \vline & $N_{\rm ROI_3}$ \vline & $N_{\rm ROI_4}$ \\
\hline
5.0 -- 5.2 \vline & 1209 \vline & 1210 \vline & 1112 \vline& 1160 \vline& 1157 \\
5.2 -- 5.4 \vline & 150 \vline & 147 \vline & 148 \vline& 150 \vline& 153 \\
5.4 -- 5.6 \vline & 51 \vline & 58 \vline & 51 \vline& 41 \vline& 43 \\
5.6 -- 5.8 \vline & 15 \vline & 13 \vline & 14 \vline& 6 \vline& 9 \\
5.8 -- 6.0 \vline & 7 \vline & 7 \vline & 2 \vline& 1 \vline& 7 \\
6.0 -- 6.2 \vline & 1 \vline & 0 \vline & 3 \vline& 1 \vline& 2 \\

\end{tabular}
\label{table:data}
\end{threeparttable}
\end{center}
\end{table}

\par The detector responses of the ROIs are obtained by tracking the ROIs through the sky and comparing with detector simulations. To handle the large ROIs and their potential nonuniform exposure, the exposure of each ROI is obtained by tracking 67 subpixels [each $\simeq (3.7 {\rm deg})^{2}$] within each ROI and then combined. We note that even though we expect the detector responses are the same for each ROI, their responses are computed separately to take into account differences in lifetime and pointing efficiencies.
The detector performance of 1/2-KM2A has been thoroughly validated with a precise measurement of the Crab Nebula~\cite{Aharonian:2020iou,Chen:2020ddg,LHAASO:2021science}. Details are presented in Ref.~\cite{Aharonian:2020iou}, and in Supplemental Material~\ref{sec:supple_flux}. For 3/4-KM2A, we use the same data selection cuts, reconstruction series, and $\gamma$/hadron separation parameters as those in 1/2-KM2A. 
Our results are subject to the same systematic uncertainties as discussed in Ref.~\cite{Aharonian:2020iou}, which is estimated to be about 7\% for the flux inference and mainly comes from the variation of event rate during the operational period due to seasonal and daily changes.
Furthermore, to assess the systematic uncertainties due to the $\gamma$/hadron separation procedure, we find that the flux would change by about 20\% if the cut condition were changed to a 30\% $\gamma$-ray survival rate. The inferred DM decay rate results are thus also subject to these uncertainties.


\textbf{Decaying dark matter formalism---}
DM decaying into various standard model states could give rise to a diffuse flux of VHE $\gamma$ rays. In the PeV energy range, the dominant $\gamma$-ray components are the prompt component generated directly from Galactic DM decays as well as the secondary component from inverse Compton (IC) scattering of electrons and positrons produced by DM particles~\cite{Cirelli:2010xx,Buch:2015iya,Esmaili:2015xpa}. Other contributions, such as bremsstrahlung and synchrotron radiation, are either subdominant or contribute at much lower energies.  Moreover, due to the cosmic $\gamma$-ray attenuation and the related electromagnetic cascade processes, extragalactic DM decays are relevant only at energies smaller than $\sim 10^4~\mathrm{GeV}$~\cite{Berezinsky:2016feh}. Thus, neglecting these contributions is conservative, and does not impact our results.

The prompt \ga-ray intensity (flux per solid angle) due to DM decay from a certain Galactic latitude~($b$) and longitude~($\ell$) is given by
\begin{equation}
    \frac{{\rm d}I^{\rm prompt}_\gamma}{{\rm d} E_\gamma} = \frac{1}{4\pi\,m_\mathrm{DM}\tau_\mathrm{DM}} \frac{{\rm d}N_\gamma}{{\rm d}E_\gamma} D(E_\gamma, b,\ell) \,,
\end{equation}
where $m_\mathrm{DM}$ and $\tau_\mathrm{DM}$ are, respectively, the mass and the lifetime of DM particles, ${\rm d} N_\gamma / {\rm d} E_\gamma$ is the photon energy spectrum per DM decay, and $D(E_\gamma, b,\ell)$ is the so-called $D$ factor. The photon energy spectrum is computed by using the \texttt{HDMSpectra} package~\cite{Bauer:2020jay}, which includes the electroweak radiative corrections. The $D$ factor is given by the integral of the DM halo density profile $\rho_h$ over the line of sight $s$, including the effect of Galactic $\gamma$-ray attenuation,
\begin{equation}
    D(E_\gamma, b,\ell) = \int_0^\infty {\rm d}s \,\rho_h\left[r(s,b,\ell)\right] e^{-\tau_{\gamma\gamma}(E_\gamma, \vec{x})} \,,
    \label{eq:Dfactor}
\end{equation}
where $\tau_{\gamma\gamma}$ is the total optical depth due to pair production ($\gamma\gamma \to e^+ e^-$) with background photons~\cite{Esmaili:2015xpa}. The photon targets are the cosmic microwave background (CMB), Galactic starlight (SL), and infrared (IR) radiation. The SL+IR background is extracted from the \texttt{GALPROPv54} code~\footnote{\url{https://galprop.stanford.edu/}.} (see also Ref.~\cite{Porter:2017vaa}). While the CMB photons are homogeneous, the SL+IR radiation depends on the position $\vec{x}$ in the Galaxy, which is expressed in terms of $(s,b,\ell)$. In particular, SL+IR dominates over CMB near the Galactic center and in the Galactic plane. Nevertheless, the angular dependence of the $D$ factor stems mainly from the DM halo density profile, for which we consider the commonly adopted Navarro-Frenk-White (NFW) distribution~\cite{Navarro:1996gj}
\begin{equation}
\rho_h(r)=\frac{\rho_s}{\left(r/r_s\right)\left(1+r/r_s\right)^2}\,,
\end{equation}
which is a function of the galactocentric radial coordinate
\begin{equation}
    r=\sqrt{s^2+R^2_\odot-2 \,s \,R_\odot \cos b\cos \ell}\,,
\end{equation}
with $R_\odot = 8.3~\mathrm{kpc}$ being the Sun position. At the scale radius $r_s=20~\rm{kpc}$ we take $\rho_s=0.33~\rm{GeV /cm^3}$, which yields a local DM density $\rho_\odot \simeq 0.4~{\rm GeV\,cm^{-3}}$~\cite{Catena:2009mf,Iocco:2015xga,Pato:2015dua}. The local DM density is found to be generally around $0.3-0.6\,\rm{GeV /cm^3}$~\cite{deSalas:2020hbh} and our DM decay results scale linearly with it. In the energy range considered, the averaged energy-dependent $D$ factor in Eq.~\eqref{eq:Dfactor} in the search region ($\mathrm{ROI}_0$) is larger by a factor ranging from 1.6 to 2.3 than those in the control regions ($\mathrm{ROI}_1-\mathrm{ROI}_4$). This ensures a higher DM intensity in $\mathrm{ROI}_0$ with respect to the other selected regions. Moreover, the DM $\gamma$-ray flux depends only slightly on the choice of the density profile for the extended DM Galactic halo and our results are robust against density profile choices, see Supplemental Material~\ref{sec:more_dm}.

The secondary Galactic IC component is computed by solving the stationary diffusion-loss equation for the electrons and positrons injected in the Galaxy by DM decays. At high energies, however, the electron-positron distribution is completely dictated by the energy losses~\cite{Esmaili:2015xpa}. Hence, by neglecting the marginal effect of diffusion, the galactic IC component takes the following expression~\cite{Cirelli:2010xx,Esmaili:2015xpa}:
\begin{gather}
\frac{{\rm d}I^{\rm IC}_\gamma}{{\rm d}E_\gamma}=\frac{1}{2\pi E_\gamma m_\mathrm{DM} \tau_\mathrm{DM}} \int_0^\infty {\rm d}s \,\rho_h(r) e^{-\tau_{\gamma\gamma}(E_\gamma, \vec{x})} \times\nonumber\\ 
\qquad\quad\int_{E_\gamma}^{\frac{m_\mathrm{DM}}{2}} {\rm d}E_e \,\frac{P_\mathrm{IC} (E_\gamma,E_e,\vec{x})}{b(E_e,\vec{x})} \int_{E_e}^{\frac{m_\mathrm{DM}}{2}}{\rm d}E'_e\, \frac{{\rm d}N_e}{{\rm d}E'_e}\,.
\label{eq:IC}
\end{gather}
Here, $P_\mathrm{IC}$ is the IC radiated power, $b(E_e,\vec{x})$ is the energy loss coefficient comprising IC and synchrotron processes, and ${\rm d}N_e / {\rm d}E'_e$ is the injected electron spectrum computed with \texttt{HDMSpectra}. For the synchrotron energy losses we adopt the regular Galactic magnetic field model with a local strength of $4.78~\mu{\rm G}$ as reported in Ref.~\cite{Strong:1998fr}. For more details on the DM signal computation and its uncertainties, see Supplemental Material~\ref{sec:more_dm}.

\begin{figure*}[t!]
\centering
\includegraphics[width=0.95\columnwidth]{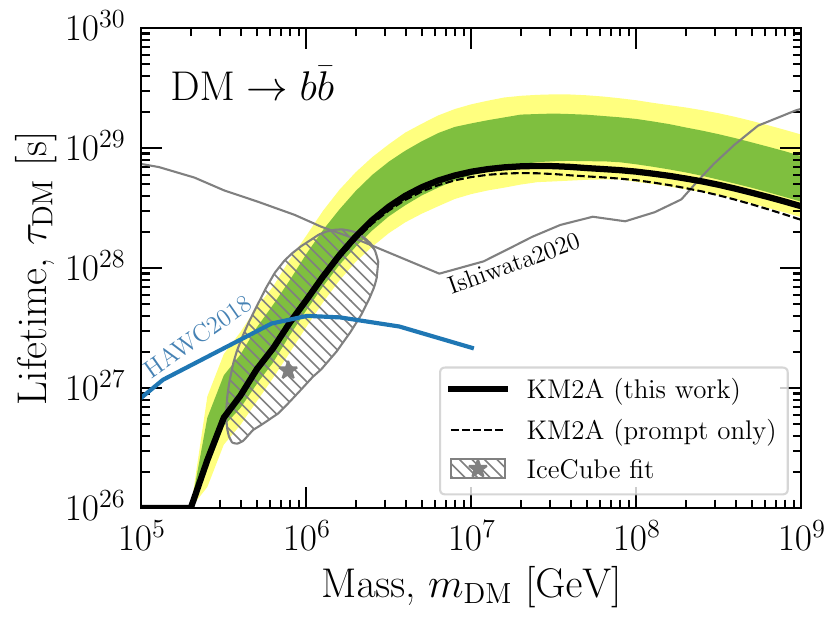}
\hspace{0.075\columnwidth}
\includegraphics[width=0.95\columnwidth]{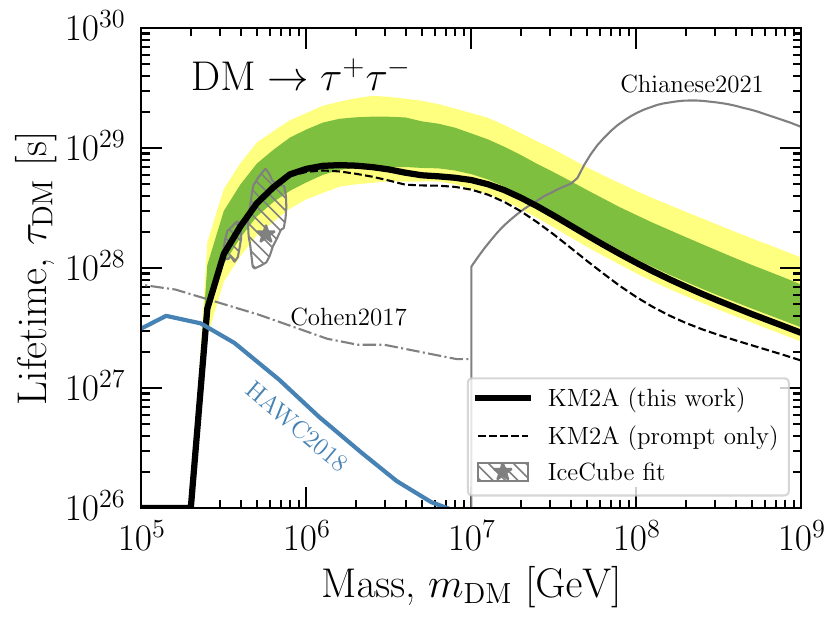}
\caption{95\% one-sided lower limits on DM lifetime obtained with the profile likelihood analysis (thick black lines), for DM decaying into $b$ quarks (left) or $\tau$ leptons (right). The black dashed line shows the limit obtained if we only consider prompt DM contribution. The green and yellow bands correspond to the expected 68\% and 95\% limit ranges from Monte Carlo simulations with the background-only hypothesis. Previous limits~\cite{Cohen:2016uyg,Ishiwata:2019aet,Chianese:2021jke} and those from HAWC~\cite{Abeysekara:2017jxs} are shown with gray and blue lines}. The hatched regions show the 1$\sigma$ DM parameter space favored by IceCube high-energy neutrino flux~\cite{Chianese:2019kyl}. 
\label{fig:constraint}
\end{figure*}

\textbf{Likelihood analysis---} 
We perform a joint-likelihood analysis on the ROIs that takes into account the DM angular distribution. The likelihood function for the $k$th ROI is given by 
\begin{equation}
    \ln {\cal L}_{k}(\tau_\mathrm{DM}, b) =  \sum_{i} N^{i}_k\ln n^{i}_{k} - n^{i}_{k}  \, , 
\end{equation}
where $N^{i}_{k}$ is the number of observed events in each energy bin, $i$, and $n^{i}_{k}$ is the modeled number of events, given by 
\begin{equation}
    n_{k}^{i}(\tau_\mathrm{DM},b)  =  \left(b^{i} + s^{i}_{k} \left(\tau_\mathrm{DM} \right)\right)  {\cal E}^{i}_{k}  \Delta\Omega \, ,
\end{equation}
where $b^{i}$ is the background model, $ s^{i}_{k}(\tau_\mathrm{DM})$ is the total integrated DM intensity for the specific ROI,
\begin{equation}
s^{i}_{k}(\tau_\mathrm{DM}) = \frac{1}{\Delta \Omega}  \int {\rm d}\Omega {\rm d}E_\gamma\, \left( \frac{{\rm d}I^{\rm prompt}_\gamma}{{\rm d}E_\gamma} + \frac{{\rm d}I^{\rm IC}_\gamma}{{\rm d}E_\gamma}  \right)  \, ,
\end{equation} 
$\mathcal{E}_k^i$ is the detector exposure on the ROI, and $\Delta\Omega$ is the solid angle of the ROIs.

Importantly, the DM intensity is different in different ROIs due to the different $D$ factor and secondary contributions, while all ROIs have the same underlying background model~($b^{i}$) due to the isotropic cosmic-ray background distribution. This breaks the signal-background degeneracy between different ROIs, and thus ${\rm ROI}_{1}- {\rm ROI}_{4}$ are included to constrain the background contribution. The background is expected to be isotropic, as the intrinsic cosmic-ray anisotropy is only $\sim$0.1\% ~\cite{icecube:aniso2013ApJ, Abeysekara:2018qho_aniso}, much smaller than the statistical uncertainties. We consider the joint-likelihood for all 5 ROIs:  $\ln {\cal L}(\tau_\mathrm{DM}, \hat{b}) = \sum \limits_{k=0}^{4}\ln {\cal L}_{k}$, with the ``hat'' signaling that the background $b^i$ has been treated as a nuisance parameter and fitted over to maximize the likelihood~\cite{Rolke:2004mj}. For the background model, $b^{i}$, we conservatively assume complete ignorance of their values in each energy bin, and thus they can take any non-negative values during the fit.

We search for the presence of a DM signal by scanning through the DM mass from $10^{5}$ to $10^{9}$~GeV for each decay channel, assuming a 100\% branching fraction. We find no significant detection of DM signals, which would correspond to a peak in the likelihood function against $\tau_{\rm DM}$. Therefore, we obtain the one-sided 95\% lower limit on the DM decay lifetime, $\tau_{\mathrm{DM, 95}}$, for each DM mass and decay channel by finding $-2\ln[{\cal L}(\tau_\mathrm{DM, 95})/{\cal \hat{L}}] = 2.71$~\cite{Cowan:2010js}, where ${\cal \hat{L}}$ is the best-fit likelihood with respect to both $\tau_{\mathrm{DM}}$ and $b$.

\textbf{Results---} Figure~\ref{fig:constraint} shows the constraints for the $\mathrm{DM}\rightarrow b\bar{b}$ and $\mathrm{DM}\rightarrow \tau^{+}\tau^{-}$ channels obtained in this work (thick black lines). Other decay channels are discussed in Supplemental Material~\ref{sec:supple_channels}. To validate our results, we perform the same joint-likelihood analysis with mock data for the ROIs using the best-fit null-hypothesis~$(\tau_{\mathrm{DM}}\rightarrow \infty)$  background model and assuming a Poisson probability distribution. The 68\% and 95\% limit ranges from such Monte Carlo simulations are shown in Fig.~\ref{fig:constraint}. We find that the actual constraints are within the 95\% expected range, but are close to the bottom range.  This is caused by a small and statistically insignificant event excess in $\rm ROI_{0}$~(The highest local significance found is about 1.4\,$\sigma$ for the $\tau^{+}\tau^{-}$ channel at $\sim 8$\,PeV.). The agreement with the Monte Carlo simulation also validates the common background hypothesis for the ROIs.  This implies that potential anisotropic astrophysical components in the ROIs, such as diffuse emission and point sources, are subdominant. In Fig.~\ref{fig:constraint} we also show the limits obtained considering only the prompt contribution to highlight the robustness of our constraints with respect to potential uncertainties in the secondary components. 

For comparison, we also show the best previous limits on DM lifetime obtained with $\gamma$ rays for both channels~\cite{Cohen:2016uyg,Ishiwata:2019aet,Chianese:2021jke}, including those from HAWC~\cite{Abeysekara:2017jxs}. Hence, the present analysis leads to a significant improvement in the DM constraints. For the $b\bar{b}$ channel, our results are about 5 times better than~\cite{Ishiwata:2019aet} around 10 PeV, while for the $\tau^+\tau^-$ channel, they are more than 10 times better than~\cite{Cohen:2016uyg} at 10\,PeV. For DM masses higher than $\mathcal{O}(10^8~\mathrm{GeV})$, our constraints are in general weaker than those obtained with KASCADE, etc,~\cite{Chianese:2021jke}.  Recently, new DM constraints~\cite{Esmaili:2021yaw, Maity:2021umk} were obtained by considering the Tibet-AS$\gamma$ data along the Galactic plane~\cite{Amenomori:2021gmk}; our constraints are generally stronger by about one order of magnitude than their model-independent limits. We emphasize that we do not consider any potential astrophysical contributions in the ROIs. Doing so will improve our constraints, but makes our results dependent on the astrophysical models.

Our limits are subject to overall systematic uncertainties, estimated to be 21\%, which is a quadrature sum of uncertainties from the detector response ($\sim$7\%) and $\gamma$/hadron separation procedure ($\sim$20\%), and mentioned above. In addition, even with the most conservative DM density profile assumption, our limit only weakens by about 5\%. These uncertainties would not affect physical interpretations of our results, as evident from Fig.~\ref{fig:constraint}.

In addition, even with the most conservative DM density profile assumption, our limit only weakens by about 5\%. These uncertainties would not affect physical interpretations of our results, as evident from Fig.~\ref{fig:constraint}.

DM decays can also produce high-energy neutrinos that can be searched for with neutrino  telescopes~\cite{IceCube:2011kcp, Esmaili:2012us, Murase:2012xs, Murase:2015gea,Aartsen:2018mxl, Bhattacharya:2019ucd,Arguelles:2019boy,Chianese:2019kyl,Dekker:2019gpe,Ng:2020ghe,Chianese:2021htv, IceCube:2022clp}. Our constraints are generally more stringent than those obtained with IceCube data~(e.g., in Refs~\cite{Chianese:2019kyl, IceCube:2022clp}), except in neutrino channels. These searches are therefore highly complementary. See Supplemental Material~\ref{sec:supple_neutrinos} for a comparison with the latest IceCube results~\cite{IceCube:2022clp}. 
Remarkably, our results highly constrain the hypothesis of decaying DM as a source of high-energy neutrinos. The limits reported in Fig.~\ref{fig:constraint} disfavor a large portion of the 68\% C.L. DM parameter space (hatched regions) and the best-fit scenario (black stars) inferred with the latest IceCube data~\cite{Chianese:2019kyl}. We note that the DM interpretation of the IceCube data is not significant ($<2\sigma$), and the IceCube data is still compatible with an isotropic spatial distribution.


\textbf{Conclusions and outlook---}
In this Letter, using 340 days of 1/2-KM2A and 230 days of 3/4-KM2A data, we obtain some of the strongest \ga-ray limits on heavy decaying DM particles. This analysis shows that, even with just partial KM2A data, LHAASO already offers unprecedented sensitivity in DM indirect-detection searches, with an immediate impact on the DM interpretation of IceCube high-energy neutrino events.

This analysis uses a data-driven method to estimate the residual cosmic-ray background through the ROIs, which allows us to obtain strong yet robust constraints on the DM lifetime. In the future, with the completion of the full KM2A array, considering more sky data and longer collection time, the effective exposure can be enhanced dramatically. Considering any underlying astrophysical components would also reduce the allowed DM contribution. Furthermore, with the full LHAASO detectors~(KM2A+WCDA+WFCTA), the \ga/hadron separation power is expected to be further improved, with the energy range extended.  Together, we expect the DM sensitivity will be significantly improved, offering new possibilities for a potential detection of DM.\vspace{1cm}

\section*{Acknowledgments} \label{sec:acknowledgements} 
The authors would like to thank all staff members who
work at the LHAASO site above 4400 meters above sea
level year-round to maintain the detector and keep the
electrical power supply and other components of the experiment
operating smoothly. We are grateful for Y.H.Yu, who cross checked the KM2A data processing. We thank Carsten Rott for helpful comments. We thank the referees for constructive comments. This work is supported in China by the National Key R\&D program of China under Grants No. 2018YFA0404201, No.2018YFA0404202, No.2018YFA0404203, and No.2018YFA0404204, by the NSFC under Grants No.U1931112, No.U1931201, No.12022502, No.11905227, No.U1831208, No.11635011, No.11761141001, No.11905240, No.11675204, No.11475190, No.U2031105, and  No.U1831129, and in Thailand by grant RTA6280002 from Thailand Science Research and Innovation. Chengdu Management Committee of Tianfu New Area provided financial support for research with LHAASO data. M.C., D.F.G.F. and G.M. acknowledge support from the research Grants No. 2017W4HA7S ``NAT-NET: Neutrino and Astroparticle Theory Network'' under the program PRIN 2017 funded by the Italian Ministero dell’Università e della Ricerca (MUR) and from the research project TAsP (Theoretical Astroparticle Physics) funded by the Istituto Nazionale di Fisica Nucleare (INFN). The work of D.F.G.F. is partially supported by the {\sc Villum Fonden} under Project No.~29388.  This project has received funding from the European Union's Horizon 2020 research and innovation program under the Marie Sklodowska-Curie grant agreement No.~847523 ‘INTERACTIONS.’ K.C.Y.N. acknowledges support by the Croucher Foundation. The work of S.A. was supported by JSPS/MEXT KAKENHI Grants No. JP17H04836, No.JP20H05850, and No.JP20H05861.

\bibliography{bib}

\clearpage
\newpage
\maketitle
\onecolumngrid

\begin{center}
\textbf{\large Supplemental Material for \\ Constraints on heavy decaying dark matter from 570 days of LHAASO observations} \\ 
\vspace{0.05in}
{LHAASO Collaboration,}\\
{S. Ando, \ M. Chianese, \ D. F.G. Fiorillo, \ G. Miele, and K. C.Y. Ng}
\end{center}
\onecolumngrid
\setcounter{equation}{0}
\setcounter{figure}{0}
\setcounter{table}{0}
\setcounter{section}{0}
\setcounter{page}{1}
\makeatletter
\renewcommand{\theequation}{S\arabic{equation}}
\renewcommand{\thefigure}{S\arabic{figure}}
\renewcommand{\thetable}{S\arabic{figure}}

The supplemental material is organized as follows. In Sec.~\ref{sec:supple_rois} we describe in detail the regions of interest selected in the present analysis. In Sec.~\ref{sec:supple_gp}, we describe the $\gamma$/hadron separation method. In Sec.~\ref{sec:supple_flux}, we detail the method used to compute the detector response from Monte Carlo simulations. In Sec.~\ref{sec:more_dm} we provide more details on the calculations of the DM signal and quantify the uncertainties related to the density profile choice, the background photons, and the galactic magnetic field. In Sec.~\ref{sec:supple_channels} we report the constraints on the lifetime of DM particles for other decay channels. Finally, we compare with the latest IceCube results in Sec.~\ref{sec:supple_neutrinos}.

\section{Regions of interest}\label{sec:supple_rois}

\begin{figure}[h!]
    \centering
    \includegraphics[width=0.45\textwidth]{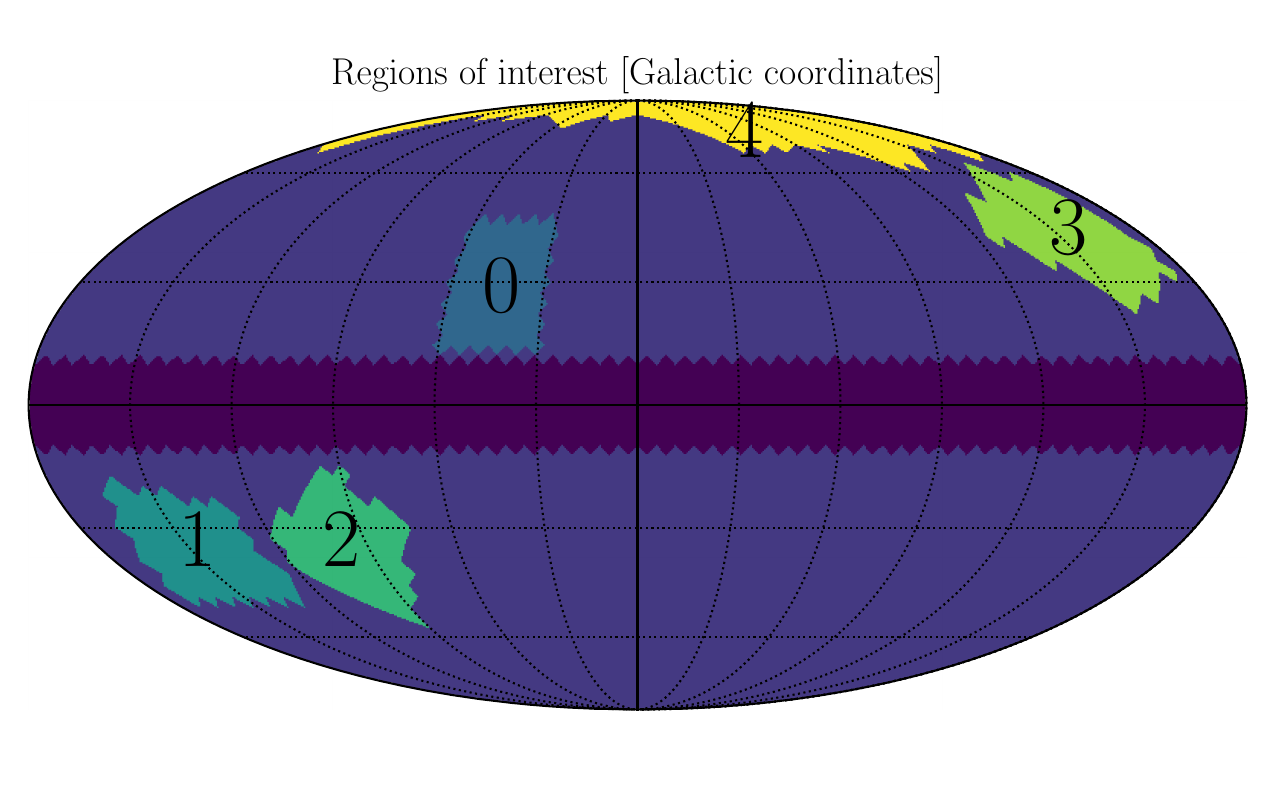}
    \includegraphics[width=0.45\textwidth]{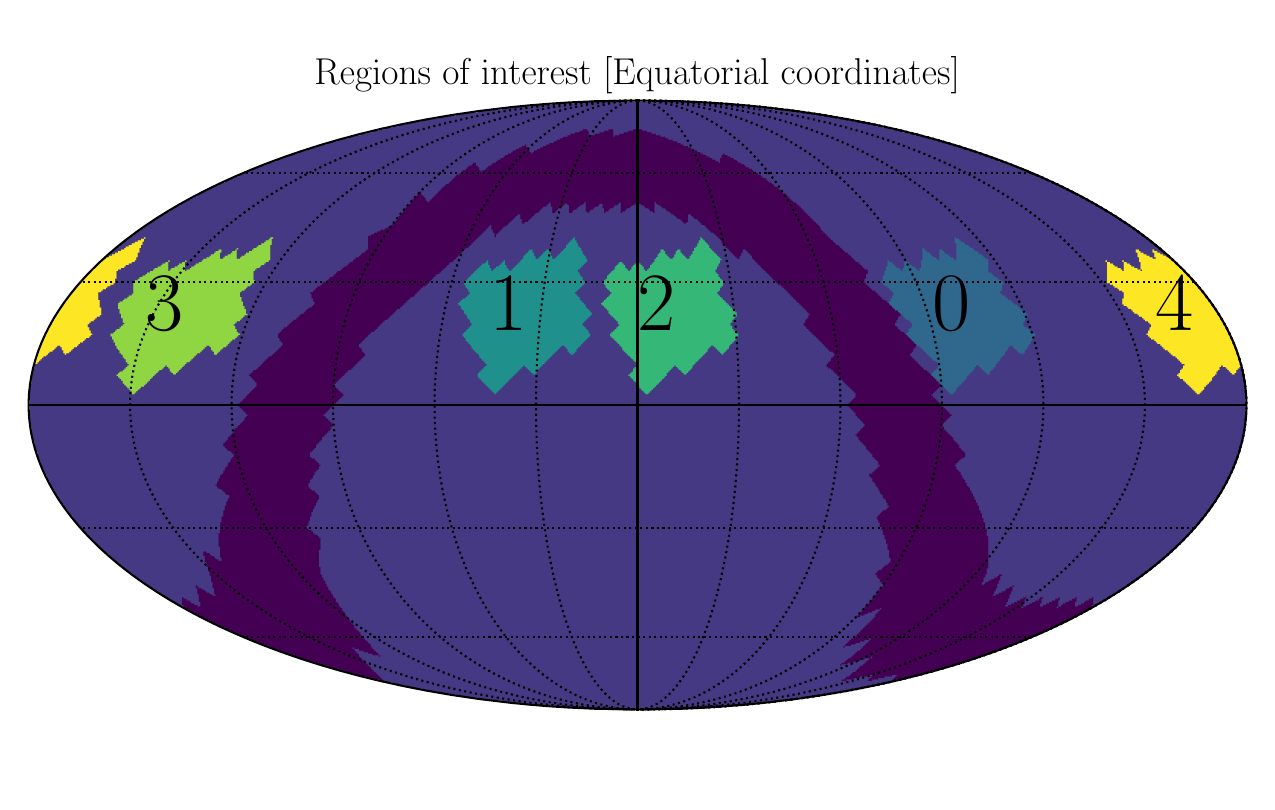}
    \caption{Pixelated sky map highlighting the selected ROIs, labeled with indices from 0 to 4, in Galactic (left panel) and equatorial (right panel) coordinates. The dark band shows 10 degrees around the Galactic plane.} 
    \label{fig:rois}
\end{figure}

In Fig.~\ref{fig:rois} we show the selected regions of interest in the pixelated sky map with Galactic (left panel) and equatorial (right panel) coordinates. In Galactic coordinates $(b,\ell$), they are centered at ($30^\circ$,$45^\circ$), ($-35.2^\circ$,$148.4^\circ$), ($-35.1^\circ$,$97.0^\circ$), ($ 43.3^\circ$,$183.3^\circ$), and ($218.1^\circ$,$77.0^\circ$).  All the ROIs are 10 degrees away from the Galactic plane (dark band in the plots) and the Fermi bubbles, in order to reduce astrophysical contamination. The search region $\mathrm{ROI}_0$ is expected to have the highest DM \ga-ray flux due to its proximity to the Galactic center. $\mathrm{ROI}_1 - \mathrm{ROI}_4$ are control regions through which the isotropic cosmic-ray background is estimated, though their DM content is also taken into account. All the ROIs have the same angular size of 0.274~sr, and have the same declination as shown in the right plot. In Tab.~\ref{tab:D_factor}, we report the energy-dependent $D$-factor, averaged over the ROIs, for the different energy bins. As the \ga-ray energy increases, the $D$-factor decreases due to the \ga-ray attenuation. Among the ROIs, the $D$-factor varies according to the NFW density profile.
{\renewcommand{\arraystretch}{1.5}%
\setlength{\tabcolsep}{10pt}%
\begin{table}[h!]
    \centering
    \begin{tabular}{c|c|c|c|c|c|c}
    \multirow{2}{*}{$\log_{10} (E_\gamma / {\rm GeV})$} & \multicolumn{5}{c|}{$D$-factor~[$10^{22}~{\rm GeV/cm^2}$]} & 
    \multicolumn{1}{c}{$N_{\rm DM}(\mathrm{ROI}_0)$} \\
    & $\mathrm{ROI}_0$ & $\mathrm{ROI}_1$ & $\mathrm{ROI}_2$ & $\mathrm{ROI}_3$ & $\mathrm{ROI}_4$ & ${\rm DM} \rightarrow b \bar{b}$, $\tau_{\rm DM} = 6.3\times 10^{28}\,{\rm s}$ \\ \hline
    5.0 -- 5.2 & 2.68 & 1.18 & 1.55 & 1.20 & 1.60 & 83.6 \\
    5.2 -- 5.4 & 2.59 & 1.13 & 1.49 & 1.15 & 1.54 & 41.9 \\
    5.4 -- 5.6 & 2.22 & 0.96 & 1.26 & 0.97 & 1.31 & 20.8 \\
    5.6 -- 5.8 & 1.66 & 0.73 & 0.95 & 0.74 & 0.98 & 6.6 \\
    5.8 -- 6.0 & 1.24 & 0.57 & 0.73 & 0.58 & 0.76 & 1.7 \\
    6.0 -- 6.2 & 1.02 & 0.49 & 0.62 & 0.50 & 0.64 & 0.4 \\
    \end{tabular}
    \caption{Dark matter $D$-factor for the NFW distribution averaged over the different ROIs for the 6 energy bins. The dependence on energy is completely due to the \ga-ray absorption.  The last column shows the expected number of detected events in the detector after $\gamma$/hadron separation for $b\bar{b}$ channel with $\tau_{\rm DM} = 6.3\times 10^{28}\,{\rm s}$ and $m_{\rm DM} = 10^7\,{\rm GeV}$. The $\tau_{\rm DM}$ value is chosen to be that of the limit in Fig.~\ref{fig:constraint}. Given the large DM mass, all energy bins receive contributions from DM decays. The values of $N_{\rm DM}$ is close to that of $2\sqrt{N_{\rm ROI_{0}}}$, showing that our limit is close to the statistical limit of the data, which highlights the importance of using the control regions as background estimates.} 
    \label{tab:D_factor}
\end{table}}



\section{ Gamma/hadron separation}\label{sec:supple_gp}

The \ga/hadron study of 1/2-KM2A has been detailed in the performance study~\cite{Aharonian:2020iou}. We follow the same procedure in this work, but with a more stringent cut for the purpose of this study, as discussed below. 

\begin{figure}[t!]
\centering
\includegraphics[width=0.5\columnwidth]{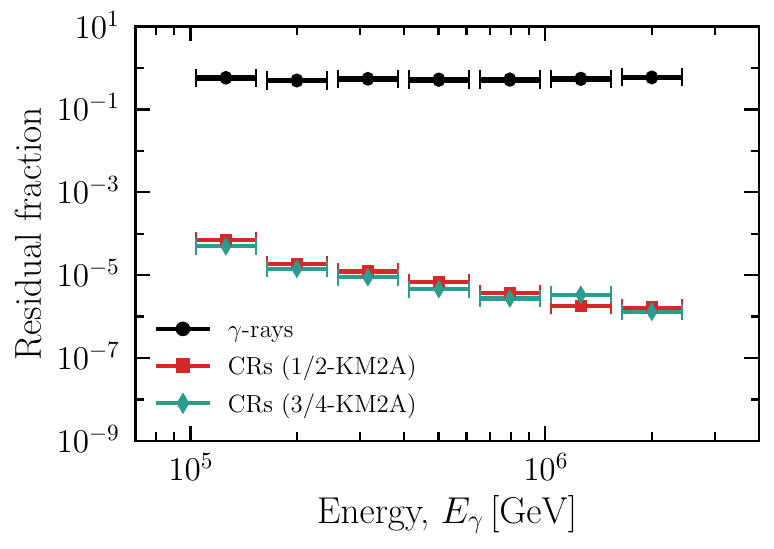}
\caption{The $\gamma$-ray and cosmic-ray survival fractions after \ga/hadron separation. \ga-ray events are obtained via simulation, while the cosmic-ray events are the actual 1/2-KM2A and 3/4-KM2A data.}
\label{fig:gp_fraction}
\end{figure}

For each triggered event, we consider the discrimination parameter $R$,
\begin{equation} 
R = \log \frac{N_{\mu}+0.0001}{N_{e}} \,,
\label{equ:cvalue}
\end{equation}
where $N_{e}$ is the number of electromagnetic particles measured by all the EDs with distance less than 200m from the shower core, and $N_{\mu}$ is the number of muons measured by all the MDs with distance less than 400m but further than 15m from the shower core. A cut at different $R$ values corresponds to different $\gamma$-ray and cosmic-ray survival efficiencies.
According to the proposed $\gamma$/hadron separation technique to identify primary cosmic rays and gamma rays by Ref.~\cite{BUGAYOV200241},  we choose the selection criteria by maximizing the quality factor Q parameter to enhance the gamma-ray observation significance in the case of background dominated sample~\cite{1983ApJ...272..317L}.

\begin{equation} 
Q=\frac{N_{\gamma,\mathrm{cut}}/N_{\gamma}}{\sqrt{ N_\mathrm{CR,cut}/N_\mathrm{CR}}} \,,
\end{equation}
where $N_{\gamma}$~($N_{\gamma,\mathrm{cut}}$) and $N_\mathrm{CR}$~($N_\mathrm{CR,cut}$) are the \ga-ray and cosmic-ray events before~(after) \ga/hadron separation. The \ga-ray survival fractions are obtained via Monte Carlo simulation, while we use the 1/2-KM2A data sets for cosmic-ray events. We also impose the condition that at least 50\% of the $\gamma$-rays are kept after the cut. With this, the survival fractions of $\gamma$-ray ($\epsilon_{\gamma}$) are slightly lower at 51.98$\%$--61.51$\%$ from 125~TeV to 2~PeV compared to Ref.~\cite{Aharonian:2020iou}, but the cosmic-ray survival fractions ($\epsilon_\mathrm{CR}$) are further lowered by about one order of magnitude to $1.86 \times 10^{-6}$ around 1~PeV, as shown in Fig.~\ref{fig:gp_fraction}. To validate our new \ga/hadron separation cut, we repeat the Crab Nebula analysis from Ref.~\cite{Aharonian:2020iou}, reproducing the Crab spectrum reported in Ref.~\cite{Aharonian:2020iou} within statistical errors. This shows that our more stringent \ga/hadron separation cut does not affect physical results. For \ga/hadron separation with 3/4-KM2A data, we take the same cut as 1/2-KM2A, and the residual fraction is also shown in Fig.~\ref{fig:gp_fraction}.

\section{Detector response}\label{sec:supple_flux}

To connect the detected number of events to physical quantities, the detector response to \ga-rays is needed, which we obtain via Monte Carlo simulation. The primary $\gamma$-ray events in Monte Carlo simulation are produced from 10~TeV to 10~PeV and sampled uniformly in logarithm space with a power-law spectrum of spectral index -2. The sampling size $N_\mathrm{mc,orig}$ varies from $10^{7}$ to $10^{5}$ as energy increases. The simulated events are sampled according to the zenith angle~($\theta$) dependence and geometrical efficiency~($\cos\theta$) of a flat detector, thus
\begin{equation}
N_\mathrm{mc,orig} \propto \sin\theta \cos\theta \, .
\end{equation}
The zenith angle is sampled from $0^{\circ}$ to $70^{\circ}$ with the azimuthal direction sampled completely. 

The ROIs are each divided into 67 identical subpixels, each with sky area about $(3.7\,{\rm deg})^{2}$. Each subpixel is then tracked across the sky fully taking into account the detector runtime corrections and efficiencies.
Consequently, the number of reference detector Monte Carlo events is given by
\begin{equation}
N_\mathrm{mc}(E_{i},\theta_{j})=
\frac{N_\mathrm{mc,trig}(E_{i},\theta_{j})}{N_\mathrm{mc,orig}(E_{i},\theta_{j})} \int_{E_{i}}\mathrm{d}E \,
\int{\mathrm{d}\Omega}\, \frac{{\rm d}I_\mathrm{mc}}{{\rm d} E}\,  S \cos\theta_{j} T(\theta_{j}) \,,
\label{equ:MCflux}
\end{equation}
where ${\rm d}I_\mathrm{mc}/{\rm d}E \propto E^{-2}$ is the reference intensity spectrum, $E_{i}$ is the energy bin considered, $\theta_{j}$ is the zenith angle from $0^{\circ}$ to $50^{\circ}$ in intervals of $1^{\circ}$, $N_\mathrm{mc,trig}(E_{i},\theta_{j})$ is the number of triggered events in the simulation~\cite{Aharonian:2020iou}, $N_\mathrm{mc,orig}(E_{i},\theta_{j})$ is the input number of events, $S$ is the reference detector area, $T(\theta_{j})$ is the time that the tracking point stays in the zenith angle considered over a day, and $\Omega_{\rm mc} = \int{{\rm d}\Omega}$ is the reference solid angle of the subpixel. The total number of reference Monte Carlo events~($N_{\rm mc}(E_{i}) = \sum_{j} N_\mathrm{mc}(E_{i},\theta_{j})$) is then summed over the zenith angles from $0^{\circ}$ to $50^{\circ}$.

Using the reference events, for each subpixel, the physical intensity~(${\rm d}I_\mathrm{data} / {\rm d} E$) corresponding to a number of detected events at each energy bin~($N_{\rm data}(E_{i})$) is then given by 
\begin{equation}
\frac{{\rm d}I_\mathrm{data}}{{\rm d} E}= \frac{N_\mathrm{data}}{N_\mathrm{mc}}\frac{\Omega_\mathrm{mc}}{\Omega_\mathrm{data}}\frac{T_\mathrm{mc}}{T_\mathrm{data}} \frac{{\rm d}I_\mathrm{mc}}{{\rm d} E}\, ,
\end{equation}
where $T_{\rm mc} = \sum_j T(\theta_{j})$, $T_\mathrm{data}$ is the actual observation time, and $\Omega_\mathrm{data}$ is the ROI solid angle. The physical intensity for each ROI is thus the averaged intensities of its subpixels.

\section{Details on the dark matter signal}\label{sec:more_dm}

\begin{figure}[t!]
\centering
\includegraphics[width=1.0\textwidth]{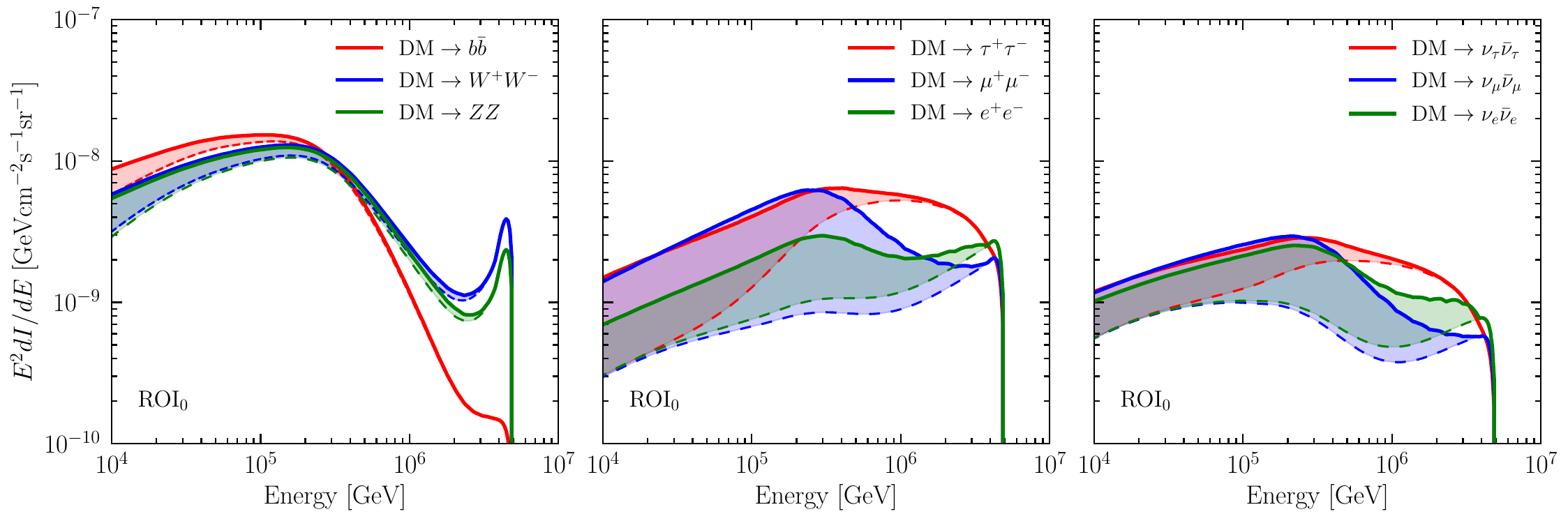}
\caption{Total DM $\gamma$-ray intensity (solid lines) from ROI$_0$ for the benchmark case of DM particles with $10^7$~GeV mass and $10^{28}$\,s lifetime. The dashed lines show the prompt Galactic component, while the shaded regions highlight the additional contribution due to the secondary Galactic IC component. Different colors correspond to different decay channels.}
\label{fig:DMfluxes}
\end{figure}

Our limits are obtained by taking into account the prompt $\gamma$-ray component as well as the secondary Inverse-Compton (IC) component from Galactic DM decays. In Fig.~\ref{fig:DMfluxes}, we show the DM $\gamma$-ray intensity from ROI$_0$ obtained in case of different DM decay channels, for the benchmark case of DM particles with mass of $10$~PeV and lifetime of $10^{28}$~s. The solid lines show the total flux given by the sum of the prompt (displayed with dashed lines) and secondary IC components. The additional contribution due to the secondary emission is highlighted by the shaded area. The fluxes stop at an energy of $m_{\rm DM}/2$, which is the maximum energy allowed in case of two-body DM decays. We emphasize that, even for neutrino channels, DM decays can produce primary $\gamma$-rays through high-order electroweak processes~\cite{Bauer:2020jay}. In the energy interval probed by LHAASO (from $10^{5.0}$ to $10^{6.2}$~GeV) the relative contribution of prompt and secondary components depends on the decay channel. As a rule of thumb, in case of DM particles decaying into quarks or electroweak bosons (left plot), the prompt component dominates over the secondary IC one for all the DM mass range considered. On the other hand, for leptophilic and neutrinophilic channels (middle and right plots, respectively) where primary electrons and positrons are produced at higher energies, the secondary IC component can dominate over the prompt one, especially in case of large DM masses. This can be seen in the figure by the size of the shaded areas.

The DM signals are affected by uncertainties that, in turn, impact our limits on the DM lifetime. In particular, both the prompt and the secondary components depend on the choice of the Galactic DM halo profile. In addition to the NFW profile, we also consider the Burkert (Bur)~\cite{Burkert:1995yz} and the Einasto (Ein)~\cite{Navarro:2003ew,Graham:2005xx} profiles:
\begin{equation}
    \rho^{\rm Bur}_h(r)=\frac{\rho^{\rm Bur}_s}{\left(1+\frac{r}{r^{\rm Bur}_s}\right)\left(1+\left(\frac{r}{r^{\rm Bur}_s}\right)^2\right)}
    \qquad {\rm and} \qquad
    \rho^{\rm Ein}_h(r)=\rho^{\rm Ein}_s \, {\rm exp}\left\{-\frac{2}{\alpha}\left[\left(\frac{r}{r^{\rm Ein}_s}\right)^\alpha-1\right]\right\}
\end{equation}
The scale radii are $r^{\rm Bur}_s=9.26~{\rm kpc}$ and $r^{\rm Ein}_s=20.0~{\rm kpc}$, respectively, while the two normalizations $\rho^{\rm Bur}_s$ and $\rho^{\rm Ein}_s$ are fixed to yield a local DM density of 0.4 ${\rm GeV \, cm^{-3}}$, in agreement with the case of NFW profile. For $b\bar{b}$ and $\tau^{+}\tau^{-}$ channels, we find that our limits change by less than 5\% in case of the Burkert and Einasto profiles. 

\begin{figure}[t!]
\centering
\centering
\includegraphics[width=0.45\textwidth]{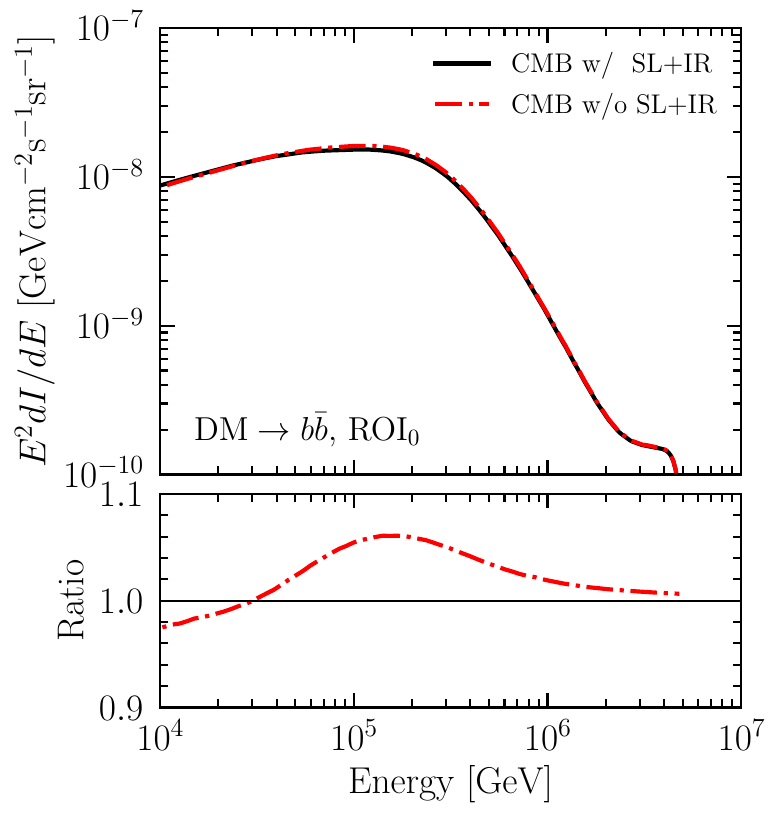}
\hspace{0.05\textwidth}
\includegraphics[width=0.45\textwidth]{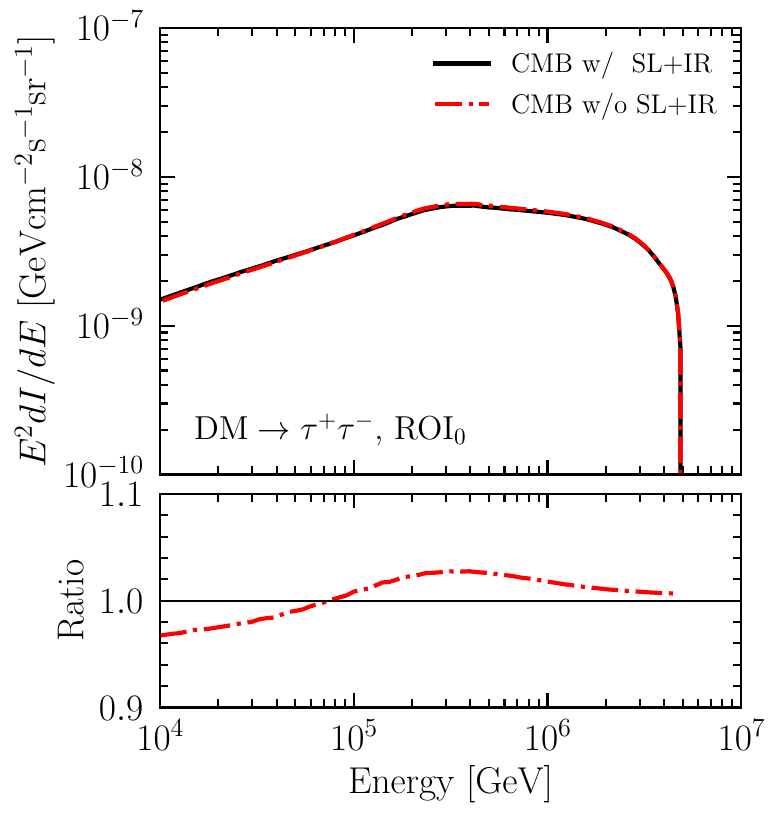}
\caption{Total $\gamma$-ray intensity from DM decays in ROI$_0$ obtained with (solid black lines) and without (dashed red lines) the SL+IR Galactic photon background, for $b\bar{b}$ and $\tau^{+}\tau{-}$ channels and with same parameters as Fig.~\ref{fig:DMfluxes}. The lower panels show the ratio of the flux without SL+IR over the that with SL+IR. }
\label{fig:noSLIR}
\end{figure}

An additional systematic uncertainty comes from the spectrum of background photons acting as targets for pair production and IC scatterings. Since the CMB is very well measured, such an uncertainty is completely related to the SL+IR background, which is typically sub-dominant with respect to the CMB.  We estimate the uncertainties associated with SL+IR background by considering the DM flux with and without the SL+IR background, as shown in Fig.~\ref{fig:noSLIR} for $b$ quark and $\tau$ lepton channels for ROI$_0$. Ignoring SL+IR has two effects. One is that the flux between $10^5$~GeV to $10^6$~GeV are \emph{enhanced} due to less $\gamma$-ray opacity in galaxy.  The second is reduced flux due to less target photons for inverse-Compton scattering, but effect is mainly seen below $10^5$~GeV, which is outside the range of our analysis.  Overall these effects only change the flux by less than 10\%.  For the other ROIs, the effect of SL+IR background is even more sub-dominant due to the very small photon number density. Hence, the uncertainty on our limits related to the SL+IR photon background can be conservatively estimated to be smaller than 10\%. We have also checked that the additional target photons from the extragalactic $\gamma$-ray background~\cite{Franceschini:2017iwq} is negligible compared to the SL+IR background in all the ROIs considered.

\begin{figure}[t!]
\centering
\centering
\includegraphics[width=0.45\textwidth]{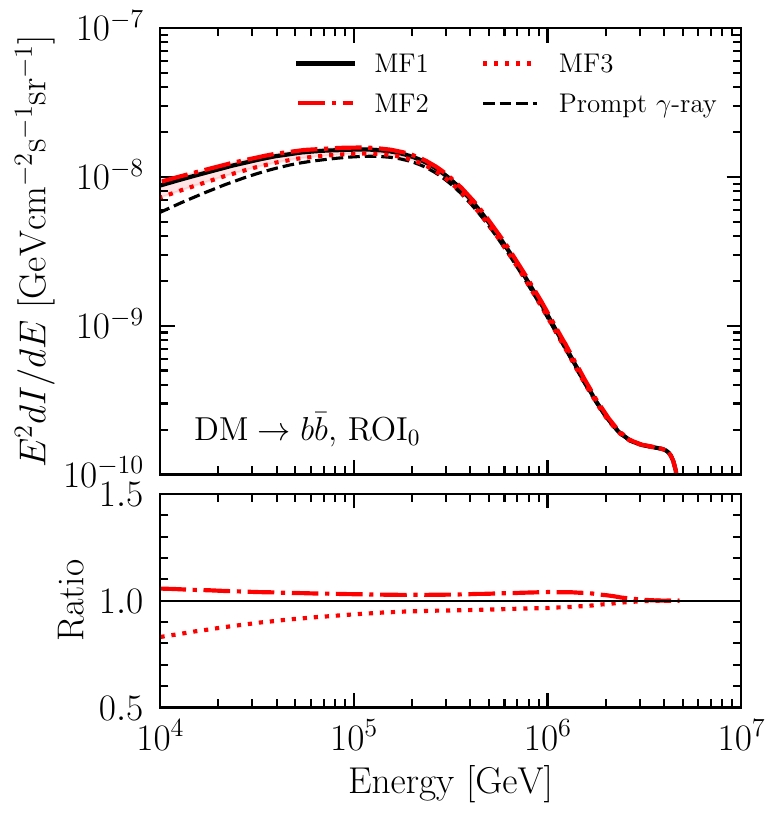}
\hspace{0.05\textwidth}
\includegraphics[width=0.45\textwidth]{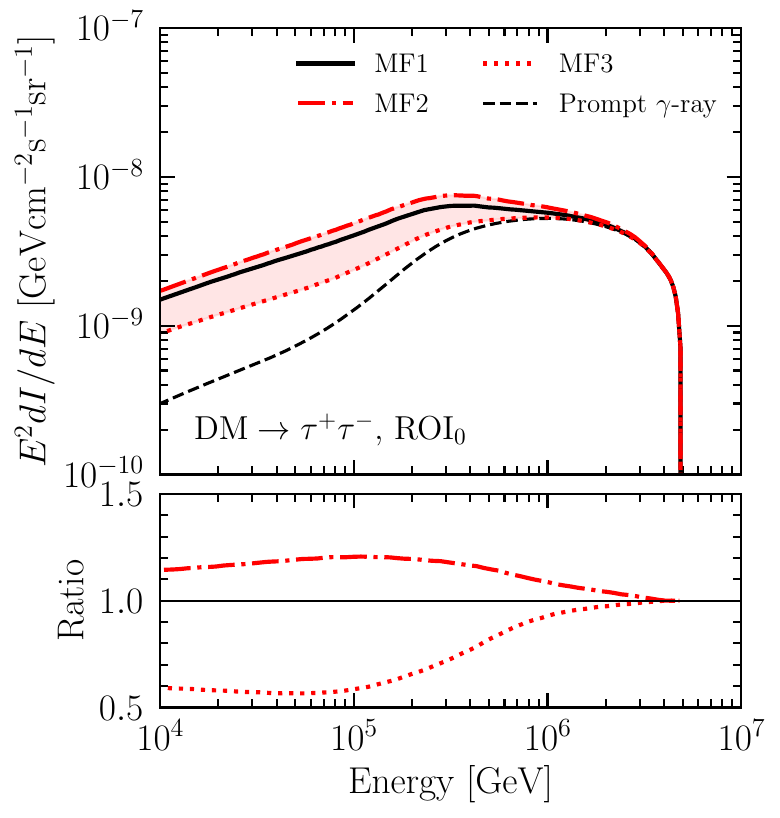}
\caption{Similar to Fig.~\ref{fig:noSLIR}, but for different MF models (see Eq.s~\eqref{eq:B} and~\eqref{eq:Bpar}), as well as the unaltered prompt $\gamma$-ray component. The lower panels show the ratio of each case over the benchmark MF1 scenario (solid black line).}
\label{fig:MF}
\end{figure}

Finally, our calculations also depend on the Galactic Magnetic Field (MF) which determines the synchrotron energy losses of electrons and positrons injected by DM particles. As such, it only affect the secondary IC component which is sub-dominant in most of the DM scenarios analyzed. In particular, the higher the magnetic field, the more efficient are the synchrotron energy losses and consequently the secondary emission is more suppressed. To estimate the uncertainty due to the Galactic MF, we follow Ref.~\cite{Buch:2015iya} and test three different configurations with the same position dependency:
\begin{equation}
    B = B_0\,{\rm exp}\left(-\frac{r-R_\odot}{r_D} - \frac{|z|}{z_D}\right)\,,
    \label{eq:B}
\end{equation}
where $r$ and $z$ are the radial distance and the height of the Galactic disk. The three MF models differ in the values assumed for the three parameters
\begin{equation}
    \begin{array}{l l l l}
    {\rm MF1:} & B_0 = 4.78~\mu{\rm G},\, & r_D = 10~{\rm kpc},\,& z_D = 2~{\rm kpc} \\
    {\rm MF2:} & B_0 = 5.1~\mu{\rm G},\, & r_D = 8.5~{\rm kpc},\,& z_D = 1~{\rm kpc} \\
    {\rm MF3:} & B_0 = 9.5~\mu{\rm G},\, & r_D = 30~{\rm kpc},\,& z_D = 4~{\rm kpc}
    \end{array} \label{eq:Bpar}
\end{equation}
The model MF1 taken from Ref.~\cite{Strong:1998fr} represents our benchmark scenario as discussed in the main text. In Fig.~\ref{fig:MF} we report the total $\gamma$-ray intensity obtained for different MF configurations, as well as the prompt $\gamma$-ray component (dashed black line). In the case of ${\rm DM} \to b\bar{b}$ (left plot), the total flux is marginally affected (less than 10\%) by the Galactic magnetic field as expected according to the sub-dominance of the secondary IC emission. On the other hand, for leptophilic and neutrinophilic channels, for which the case ${\rm DM} \to \tau+\tau^-$ (right plot) is representative, the total flux can vary up to $\sim 50\%$. Nonetheless, in all the cases, the total flux including the secondary emission is always higher than the prompt component. Therefore, the limits obtained using the prompt component alone are conservative and robust.

\section{Additional dark matter decay channels}\label{sec:supple_channels}

\begin{figure}[t!]
\centering
\includegraphics[width=0.95\textwidth]{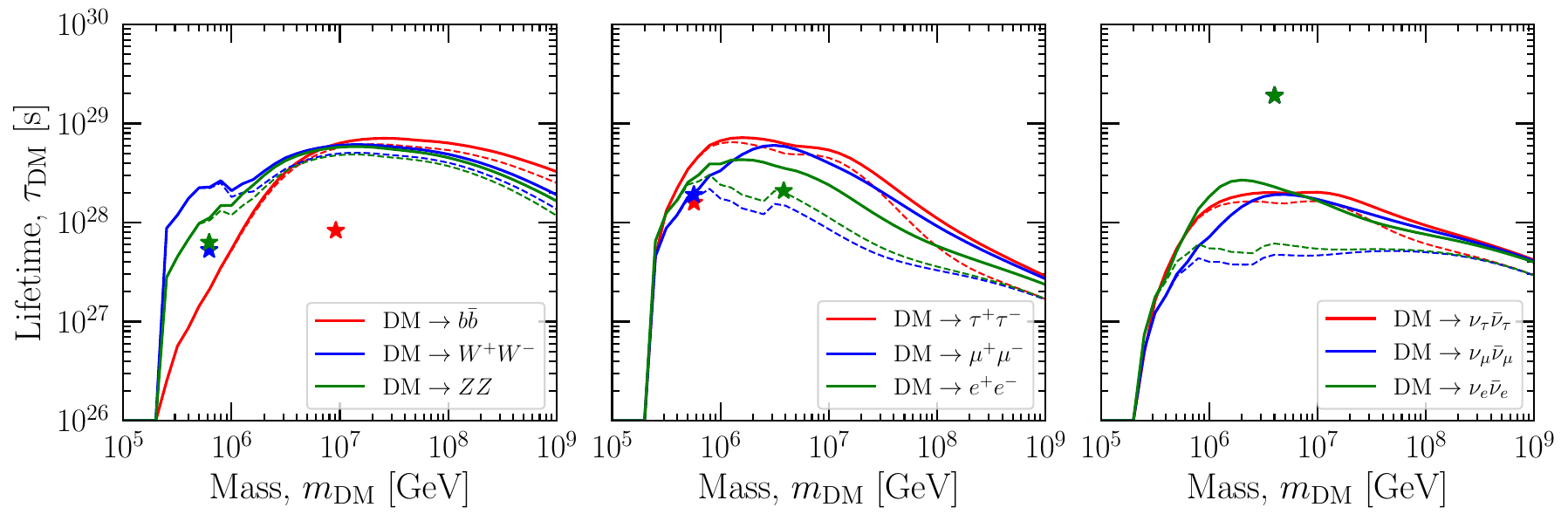}
\caption{Constraints at 95\% CL on DM lifetime obtained with the profile likelihood analysis for the DM decay channels. The lines of the $b\bar{b}$ and $\tau^+\tau^-$ constraints corresponds to the ones reported in Fig.~1. The dashed lines correspond to the limits obtained considering only the prompt contribution from DM decay. The stars correspond to the best-fit scenario from the 7.5-year IceCube HESE data~\cite{Chianese:2019kyl}.}
\label{fig:constraint_all}
\end{figure}

In Fig.~\ref{fig:constraint_all} we report the constraints on the DM lifetime obtained with the profile likelihood analysis for different DM decay channels. The limits displayed with solid lines are derived by including both prompt and secondary Galactic photons from decaying DM particles. The limits shown with dashed lines are obtained by taking into account only the prompt emission from DM decays and, therefore, are robust against the uncertainties described in the previous section. In plots we also show the DM best-fit points (stars) obtained by analyzing the 7.5-year IceCube HESE data only under the DM interpretation of the diffuse high-energy neutrino flux~\cite{Chianese:2019kyl}. 
The best-fit points differ from those reported in Fig. 1 (in the main text), which have also taken into account the 10-year IceCube through-going muon neutrino data (see Ref.~\cite{Chianese:2019kyl} for more details). As can be seen from the figure, our results are in tension with the hypothesis of DM contributions in the diffuse high-energy neutrino flux for all the decay channels except for the neutrinophilic ones (right panel).

\section{Comparison with IceCube}\label{sec:supple_neutrinos}

\begin{figure}[t!]
\centering
\includegraphics[width=0.5\columnwidth]{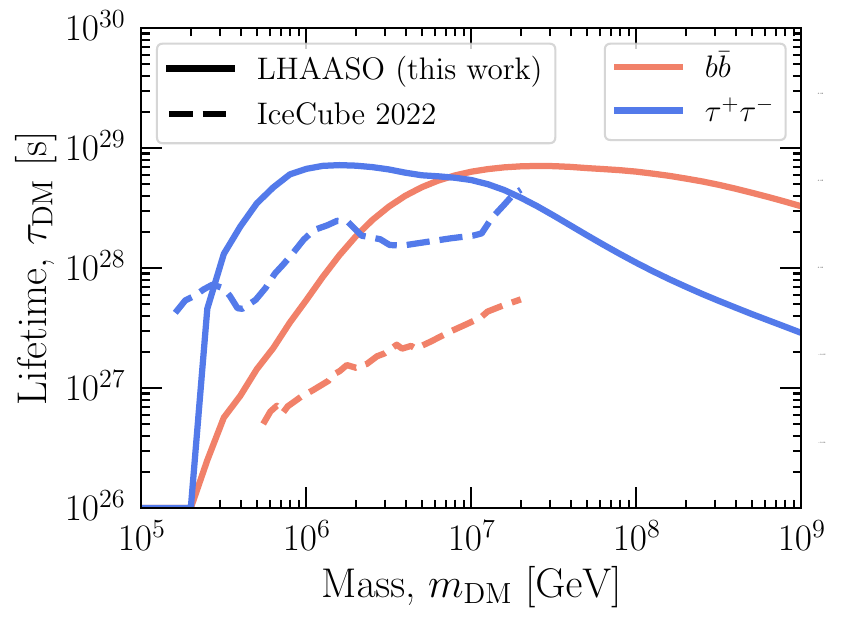}
\caption{Solid lines show the limits obtained in this work with LHAASO for $b\bar{b}$ (orange) and $\tau^{+}\tau^{-}$ (blue) channels, respectively.  For comparison, in dashed lines, we show the limits from the most recent IceCube results~\cite{IceCube:2022clp}.  }
\label{fig:neutrino}
\end{figure}

Figure.~\ref{fig:neutrino} shows the comparison of our limits with the latest IceCube results from Ref.~\cite{IceCube:2022clp}. In the two channels considered ($b\bar{b}$ and $\tau^{+}\tau^{-}$), our results are generally stronger.  However, IceCube would be stronger in channels involving neutrino finals states, which highlights the complementarity between gamma-ray and neutrino detectors. 

\end{document}